      [1999/12/01 v1.4c Il Nuovo Cimento]
\documentclass{cimento}

\usepackage{graphicx}  
\title{Kiso observations of 20 GRBs in the HETE-2 era}
\author{Y.~Urata\from{ins:a}\from{ins:b}\ETC,
Y.~Nakada\from{ins:c}, 
T.~Miyata\from{ins:c}, 
T.~Aoki\from{ins:c}, 
T.~Soyano\from{ins:c}, 
K.~Tarusawa\from{ins:c}, 
H.~Mito\from{ins:c},
S.~Nishiura\from{ins:c}, 
T.~Tamagawa\from{ins:a}\atque
K.~Makishima\from{ins:a}\from{ins:e}
}
\instlist{
\inst{ins:a} RIKEN (Institute of Physical and Chemical Research), 2-1 Hirosawa, Wako, Saitama 351-0198, Japan
\inst{ins:b} Department of Physics, Tokyo Institute of Technology, 2-12-1 Oookayama, Meguro-ku, Tokyo 152-8551, Japan
\inst{ins:c} Kiso Observatory, Institute of Astronomy, University of Tokyo, Mitake-mura, Kiso-gun, Nagano 397-0101, Japan
\inst{ins:e} Department of Physics, The University of Tokyo, 7-3-1 Hongo, Bunkyo-ku,Tokyo 113-0033, Japan
}
\PACSes{\PACSit{95.55.Cs}{Ground-based ultraviolet, optical and infrared telescopes} 
\PACSit{98.70.Rz}{ gamma-ray source; gamma-ray bursts}}
\begin{document}

\maketitle

\begin{abstract}

We established a GRB follow-up observation system at the Kiso
observatory (Japan) in 2001.  Since prior to this the east Asian area
had been a blank for in GRB follow-up observational network, this
observational system is very important for studying the temporal and
spectral evolution of early afterglows. We have thus been able to make
quick observations of early phase optical afterglows based on {\it
HETE-2} and {\it INTEGRAL} alerts.  Thanks to this quick follow-up
observational system, we have so far been able to use the Kiso
observatory to monitor 20 events, and conduct follow-up observations
in the optical and near infrared wavelengths.

\end{abstract}

\vspace{-0.2cm}
\section{Kiso Observatory}
\vspace{-0.2cm}

The TOO system for prompt GRB follow-up observations has been in
operation at the Kiso observatory since 2001 \cite{urata04a}. We have
also established a follow-up observation system based on this system
at the Lulin observatory, Taiwan \cite{huang}. The Kiso observatory is
located in Nagano-Prefecture, Japan, It has a 105 cm Schmidt telescope
and two other instruments, a 2k$\times$2k CCD camera, and a near
infrared camera named KONIC (Kiso Observatory Near-Infrared
Camera). The 2k$\times$2k camera's FOV. is 50 $\times$50 arcmin and
the limited magnitudes are 22.0 mag, 22.5 mag, 21.0 mag and 21.0
mag. (for each B,V,R,I band, 10 $\sigma$, 900 sec exposure). The
KONIC's FOV. is 20 arcmin and the limited magnitude are 13.0 mag, 12.3
mag and 10.8 mag (for each J,H,K band). Otherwise, we also use two
objective prisms. These prisms allow low-dispersion slit less
spectroscopy with a 2k $\times$ 2k CCD. The FOV is 50 $\times$50
arcmin.

\vspace{-0.3cm}
\section{Observations}
\vspace{-0.2cm}
 We have thus made 20 follow-up observations in response to the {\it
HETE-2} and {\it INTEGRAL} alerts , and successfully detected the
optical afterglows of 6 GRBs.  There 20 observations (16 GRBs and 4
XRFs) were performed earlier than the delay timing ($\sim8$ hours) of
the {\it BeppoSAX} position alerts.  In the {\it BeppoSAX} era, the
afterglows were only observable for a single power-law, or at best,
the jetted achromatic broken power-law behavior could be seen.

\begin{figure}
\includegraphics[width=0.5\textwidth]{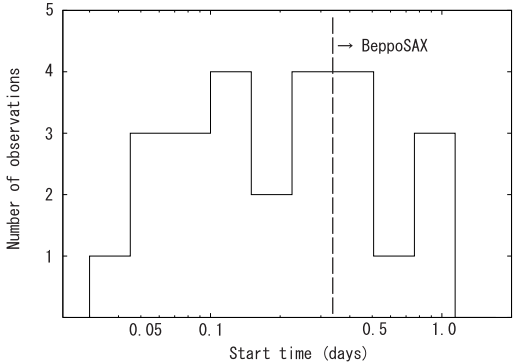}     
\caption{The number of follow-ups, plotted against the time 
when the observations were started. The dashed line indicates the
typical delay time of the {\it BeppoSAX} alerts($\sim8$ hours) .}
\end{figure}

\vspace{-0.3cm}
\section{Results}
\vspace{-0.1cm}
\subsection{GRB 030329}

In figure 1, we plot the $Rc$ band light curve of the afterglow of GRB
030329, based on our photometry, using the Kiso and RTT150 data. Over
a time range of 0.075 - 0.293 days, the light curve is well fitted for
a single power law, which temporal index is $\alpha=-0.894\pm0.004$
with a reduced $\chi^{2}$($\chi^{2}/\nu$) of 1.12 for $\nu=53$.  In
order to better constrain the early ($<$ 0.3 days) behavior of the
light curve, we combined our data with the two $Rc$ band photometric
points observed using the SSO 40-inch telescope and reported in Price
et al. (2003); $Rc$=12.6$\pm$0.015 at 0.065 days and
$Rc$=12.786$\pm$0.017 at 0.073 days. These observations are among the
earliest filtered observations of this afterglow.  We have
successfully fitted the combined $Rc$ band light curve again to a
single power law, form which the decay index is $-0.891\pm0.003$, with
$\chi^{2}/\nu$=0.817, for $\nu=36$.  Thus, we have successfully fitted
the very early (0.065 - 0.293 days) $Rc$ band light curve to a single
power law.
Obviously, our light curve does not show the wriggle structure
reported by Uemura et al (2003).  For a more quantitative comparison,
we tried to fit the present light curve (Kiso, SSO and RTT150) to the
best-fit model function reported by Uemura et al. (2003), in which the
all over normalization alone is allowed to vary.  The fit failed with
$\chi^2/\nu=6.91$ for $\nu=50$.  Based on this large value of
$\chi^2/\nu$, we can rule out the Uemura et al.'s model (2003), at a
more than 99.99\% confidence level.

There are two possible explanations for the structures reported by
Uemura et al (2003), and Sato et al (2003). First, Uemura et al
included a variable star in their list of standard stars, and used it
in their calibrations.  Second, an ``under-sampling system'' (which
means that the pixel scale for the sky is larger than the point spread
function of the sources), due to using a ``front illuminated CCD'',
which potentially produce a wiggling light curve, even if we observe
stable objects.  This is because the electrode structure on the CCD
chip reduces the brightness of as object located on it. When
observations are performed with de-focus, or larger sized seeing
conditions, this effect is relaxed
\cite{oshima}.
They however do not discuss the information of their system clearly,
 such quick GRB observations system tend to lead to ``under-sampling''
 when entire GRB error regions are concerned.

\begin{figure}
\includegraphics[width=0.45\textwidth]{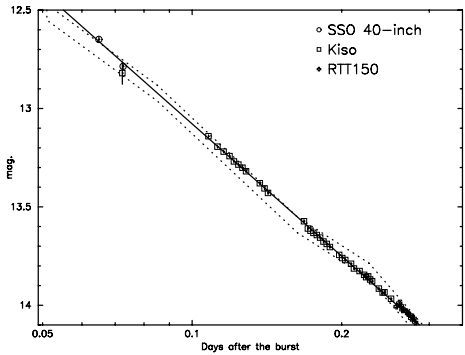}
\caption{The $Rc$-band light curve based on the photometry at Kiso,
shown together with the SSO (Price et al. 2003) and the RTT150
(Burenin et al. 2003) data points. The solid line indicates the best
fit power-law to the Kiso, RTT150 and SSO points. The dashed lines
indicate $\pm$0.04 mag error band around the unfiltered light curve
reported by Uemura et al (2003).}
\end{figure}

\vspace{-0.1cm}
\subsection{GRB 021004}

We started follow-up observations of the GRB 021004 field with a 1.05
m Schmidt telescope at 12:06:13 UT on 2002 October 4 (80 min after the
burst). The early afterglow spectrum was first detected utilizing the
strong point of the slit-less spectrometes before optical
identification.
The spectrophoto metric results are plotted together with the Bisei
$Rc$ band points \cite{kawabata} as seen in figure 3 (left).  The
$Ic$, $Rc$ and $Ic$ band light curves early show the re-brightening
phase. The photometric points are fitted to smooth broken power law
function. The re-brightening is independent of the color. Its indices
are $\sim-2$.
These characteristics are different from the predictions of the
standard afterglow model. The temporal index $\alpha_{1}\sim2$ is
steeper than that of the value ($\sim0.5$) predicted by the standard
afterglow model.
The reddening maps of Schlegel et al.(1998) give a Galactic reddening
of $E_{B-V}=0.060\pm0.020$ mag in the optical afterglow direction. The
corresponding Galactic extinction are $A_{V}=0.195$, $A_{Rc}=0.160$,
and $A_{Ic}=0.116$. The spectral index at 0.142 days after the burst
is $\beta=-1.01\pm0.10$.
Although, the light curves are not consistent with the standard
afterglow model for $<0.1$ days, the $\beta$ at 0.142 days is close to
value predicted by the model, in the regime of $\nu_{c}<\nu_{opt}$,
with a spherical geometry.

\begin{figure}
\includegraphics[width=0.5\textwidth]{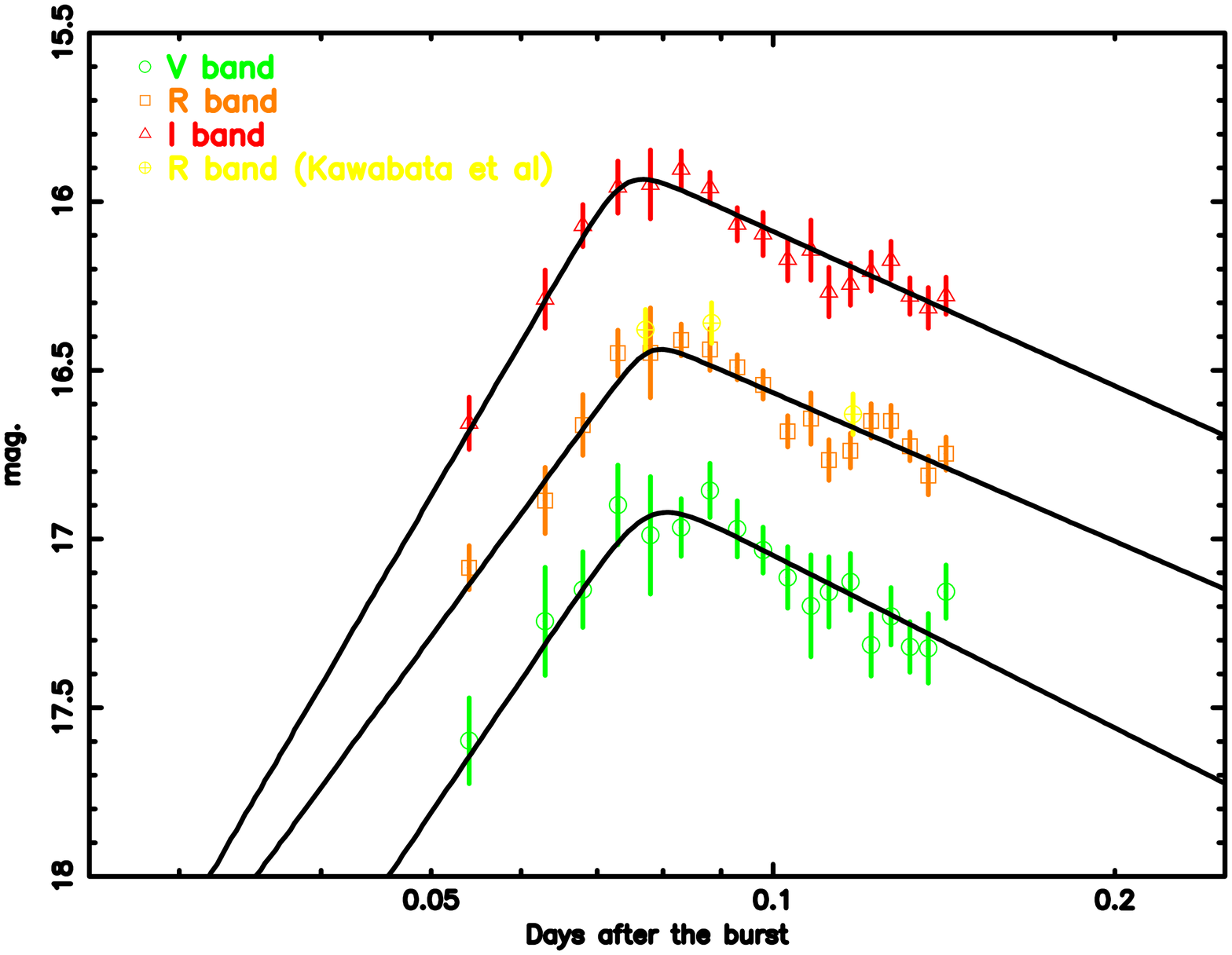}
\includegraphics[width=0.5\textwidth]{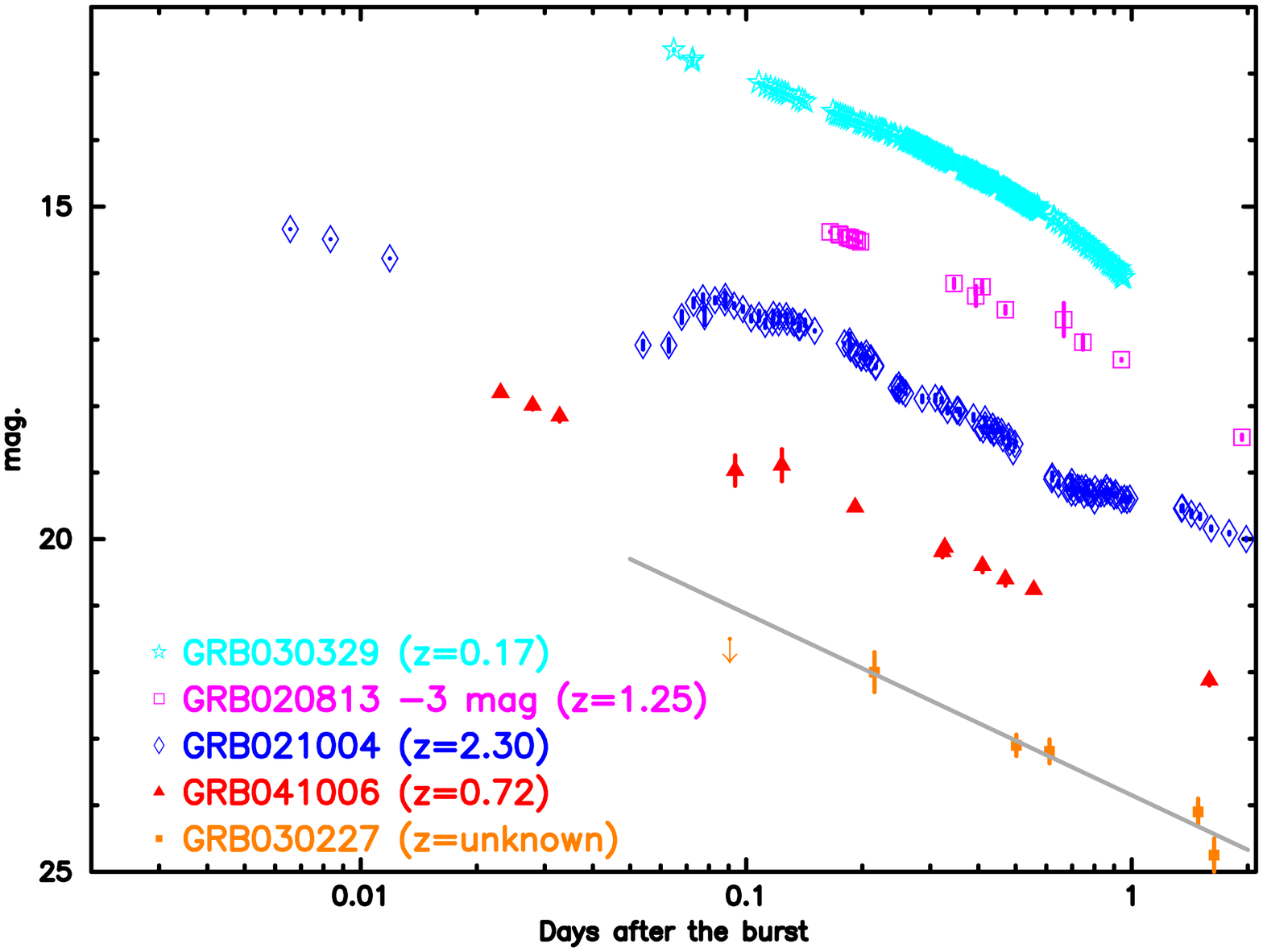}
\caption{(left) The $V$, $Rc$ and $Ic$ band light curve of GRB 021004. The solid lines indicate the best fit power-law functions. (right) The $Rc$ band light
curves of the 5 GRB afterglows which we observed in the early phases.}
\end{figure}

\vspace{-0.1cm}
\subsection{GRB 030227}

In figure 3 (right), we plot the upper limit together with the $Rc$
band photometric points reported by Castro-Tirado et al. (2003). The
single power law function which has an index of $\alpha=-1.10\pm0.14$
with $\chi^{2}/\nu=1.53$ (Castro-Tirado et al. 2003) is also shown.
Over a period of 0.09$\sim$0.1 days after the burst, our upper limits
give stronger constraints than the back-extrapolated power-law
evolution line.  These results imply that there is a plateau phase
around 0.1 days after the burst, as found for GRB 041006 (Urata et
al. 2005). According to the standard afterglow model, the afterglow
component is expected to show a brightening phase with a temporal
index about 0.5. Thus, the break time is constrained as $0.12\sim0.20$
days after the burst, and the upper limit of the peak calculated
brightness is $Rc=21.3$. The former value is close to that of GRB
041006 in the $Rc$ band, $\sim0.1$ days.

As described, the early light curves of 4 out of 5 afterglows (figure
3 right), GRB~020813 (Urata et al. 2003a), GRB~021004 (Urata et
al. 2003b), GRB~041006 (Urata et al. 2005) and GRB~030227, also
deviate significantly from the single power-law decay.  Such effects
are particularly prominent in the cases of GRB~041006, GRB~030227 and
GRB~021004.  In fact, the $Rc$ band light curve of the GRB~041006
afterglow exhibits a plateau phase around 0.09 days (Urata et
al. 2005).  The afterglow of GRB 021004 shows a clear re-brightening
phase which peak at $\sim 0.07$ days (Urata et al. 2003b) and that of
GRB~030227 is suggestive of a similar plateau at $0.12\sim0.20$ days
after the burst.
These results, based mainly on our own data, also indicate that the
early optical afterglow emissions are composed of two components: the
initial ``flash'' or other jet component emissions which have a fast
power-law decline and the genuine afterglow component, which emerges
after of a few hours.

\vspace{-0.2cm}
\acknowledgments
\vspace{-0.2cm}
We thank the staff and observers at the Kiso observatory for the
various arrangements. Y.U.was wish to acknowledges support of the
Japan Society for the Promotion of Science (JSPS) through a JSPS
Research Fellowship for Young Scientists.

\end{document}